\documentclass[10pt,prl,aps,twocolumn,showpacs,floatfix]{revtex4}
\usepackage{graphicx}
\usepackage{amsmath}
\usepackage{epstopdf}
\usepackage{subfigure}
\usepackage{float}
\usepackage{amsbsy}
\usepackage{color}

\begin{document}
\draft
\title{Violation of the Spin Statistics Theorem and the 
Bose-Einstein Condensation of Particles  with Half Integer Spin}
\author{H. D. Scammell and O. P. Sushkov }
\affiliation{School of Physics, The University of New South Wales,
  Sydney, NSW 2052, Australia}
\date{\today}
\begin{abstract}
We consider the Bose condensation of particles with spin 1/2.
The condensation is driven by an external magnetic field.
Our work is motivated by ideas of quantum critical deconfinement
and bosonic spinons in spin liquid states.
%by results of Quantum Monte Carlo simulations revealing
%properties of deconfined bosonic spinons.
We show that both the nature of the novel Bose condensate and the excitation 
spectrum are fundamentally different from that in the usual integer spin case.
We predict two massive (``Higgs'') excitations and two massless
Goldstone excitations. One of the Goldstone excitations has a linear
excitation spectrum and another has quadratic spectrum. This implies that 
the Bose condensate does not support superfluidity, the Landau criterion is 
essentially violated.
We formulate a ``smoking gun" criterion for searches of the novel Bose condensation.

\end{abstract}
\pacs{64.70.Tg%(Quantum phase transitions)
, 75.40.Gb% (Critical-point effects- Dynamic properties)
, 75.10.Jm%(Quantized spin models)
%, 74.20.De% ? Phenomenological theories (two-fluid, Ginzburg-Landau, etc.)?
%, 64.60.De% ? Statistical mechanics of model systems (Ising model, Potts model, field-theory models, Monte Carlo techniques, etc.)
%, 64.60.Cn% ? Order-disorder transformations 
%, 75.10.Dg% ? Crystal-field theory and spin Hamiltonians
}

\maketitle
\newpage

The phenomenon of Bose-Einstein condensation~\cite{Bose,Einstein} is of enormous importance in physics,
from superfluidity of liquid $^4$He~\cite{London,Landau,Bogoliubov} to cold atoms~\cite{Anderson}.
The famous spin-statistics theorem~\cite{Pauli} claims that particles with integer spin obey Bose
statistics and particles with half integer spin obey Fermi statistics.
Therefore only particles with integer spin can Bose condensate~\cite{com1}. 
The spin statistic relation is based on 
Lorentz invariance,  so the relation is a relativistic effect which is absolutely important even at
extremely low energy.
If the number of particles is conserved the stability of Bose condensate is not related to interaction between
bosons. The repulsive interaction is only responsible for a linear in momentum spectrum of Goldstone
excitations~\cite{Bogoliubov}. The linear spectrum guarantees superfluidity of the Bose 
condensate~\cite{Landau}.
If the number of particles is not conserved, like in the case of magnetic field induced Bose condensation of 
magnons in a gapped quantum antiferromagnet~\cite{Nikuni,Giamarchi}, then the stability of the 
condensate is only due to repulsion between particles. The excitation spectrum above the 
condensate is still linear in  momentum. The linearity is a generic property of a conventional
Bose condensation.

The idea of bosonic quasiparticles with spin 1/2, spinons, is a new paradigm in
condensed matter physics. Bosonic spinons have been proposed as quasiparticles in some kinds
of spin liquids~\cite{Wen2002}. Bosonic spinons also appear in the
deconfined quantum criticality (DQC) scenario. 
The scenario has been suggested~\cite{Senthil,SenthilA} to explain
properties of a magnetic quantum phase transition between a Neel antiferromagnet (AF) and a 
valence-bond solid (VBS) in a two-dimensional system of S=1/2 spins.
According to this scenario the system fractionalize into independent (deconfined)
spinons. 
Bosonic spinons do not contradict the spin statistics theorem since they are quasiparticles in a 
solid and hence the Lorentz invariance is explicitly violated.
Both the spin liquid and the DQC are essentially strong coupling nonperturbative phenomena, 
therefore there is no simple theoretical model where the phenomena can be studied analytically.
There are several spin-lattice models which, according to numerical studies,
manifest the spin liquid behaviour or DQC. 
The ‘‘J-Q’’ spin-lattice model had been specially designed~\cite{Sandvik07}
to study DQC numerically. The model is assessable by  quantum Monte Carlo (QMC) numerical 
method which allows the study of very large lattices. To the best of our knowledge
this model enables the analysis of the largest lattices among all existing DQC and spin liquid lattice models.
Therefore, we refer to numerical results on this model.

The QMC simulations of the ‘‘J-Q’’ model~\cite{Sandvik07,Melko} support the DQC scenario.
The scenario violates the long-held ‘‘Landau rule’’ according to which an order-order
transition breaking unrelated symmetries should be of first order. Some studies claim a generic 
first-order Neel-valence-bond-solid transition~\cite{Kuklov}.
A weak discontinuity (first order) certainly cannot be ruled out based on numerical data.
However, 
 the most important issue is not the order of the transition, the most important issue
is deconfinement of spinons. This issue was directly addressed in Ref.~\cite{Sandvik11}
using the Wilson ratio method (the ratio of uniform magnetic susceptibility over specific heat). 
The ratio was calculated  numerically (QMC) within the strongly 
interacting J-Q model. The calculation has been performed in terms of original (bare)
lattice spins
%objects of the model 
and does not refer to any quasiparticles. On the other hand the
Wilson ratio can be expressed in terms of quasiparticles, importantly the ratio is proportional to 
the  quasiparticle spin squared. Comparison of the QMC results with the quasiparticle
expression has demonstrated the following points.
(i) The spin of the bosonic quasiparticle is really  S=1/2.
This is a major conclusion. Besides that the analysis~\cite{Sandvik11} has 
demonstrated two more points. (ii) The spinons interact even weaker
than it is predicted in the DQC scenario~\cite{Senthil,SenthilA}. (iii) There is a logarithmic 
correction to the Wilson ratio, the correction is due to large spinon occupation numbers
in a weak magnetic field, $n_k\gg 1$, and hence, the correction is a precursor to spinon Bose condensation.

In the present work we consider Bose condensation of particles with S=1/2.
The condensation is  driven by an external magnetic field.
The present study is motivated by models of spin liquids~\cite{Wen2002},
the DQC scenario~\cite{Senthil,SenthilA} and by numerical results
of Ref.~\cite{Sandvik11} for the J-Q model.
However, results of the present work are decoupled from the motivations. The results are generic and can be 
applied to bosonic spinons in solids and to particles in vacuum which for some reason do not 
respect the  spin-statistics theorem (the case of fundamental violation of Lorentz invariance).

Following Refs.~\cite{Senthil,SenthilA} we use the CP$^1$ representation to describe spinons,
so the mathematical object of the theory, $z$, is an SU(2) spinor. The effective Lagrangian of spinons is
\begin{eqnarray}
\label{L}
\mathcal{L}&=&\{\partial_t z^\dag + iSz^\dag(\vec{\sigma}\cdot\vec{B})\}\{\partial_t z - iS(\vec{\sigma}\cdot\vec{B})z\}\nonumber\\
 &-&(\nabla z^\dag)(\nabla z) -m^2z^\dag z - \frac{\alpha}{2}(z^\dag z)^2.
\end{eqnarray}
We set the spinon speed equal to unity,
${\vec{\cal B}}=2S\mu_B\vec{B}$, $ \vec{B}$ is the external 
uniform static magnetic field, $S=1/2$ is spin of the spinon,
and $\mu_B$ is the Bohr magneton;
${\vec {\sigma}}$ are the usual Pauli matrices, 
$m$ is mass of the spinon.
The interaction is repulsive, $\alpha > 0$ and first  we assume that 
$m^2 \ge 0$.  
Unlike Refs.~\cite{Senthil,SenthilA} the Lagrangian (\ref{L}) 
does not contain a dynamic gauge field. This is directly motivated by the QMC 
analysis~\cite{Sandvik11} which does not show a contribution of the gauge field.
We already pointed out that the same QMC analysis indicates a logarithmic correction 
to magnetic susceptibility. The correction can be
due to a slightly nonpolinomial interaction of spinons.
Here we disregard this correction and consider a simple quartic interaction
$\frac{\alpha}{2}(z^\dag z)^2$.
% \to \frac{\alpha}{2}(z^\dag z)^{d}$, $d\approx 1.9$.
%Here we disregard the small deviation from the .
% since the deviation does not influence  the Bose condensation properties.
While the DQC motivation comes from  two-dimensional (2D) systems (spatial 
dimensions),  the Lagrangian (\ref{L}) can be considered both in two and three dimensions.

If magnetic field is smaller than mass, ${\cal B} < m$, then the classical part
of the field $z$ is zero and the quantum part obeys the following Euler-Lagrange equation
\begin{eqnarray}
\label{EL}
\ddot{z}- \nabla^2 z -2i(\vec{\sigma}\cdot\vec{{\cal B}})\dot{z} + 
[m^2-{\cal B}^2]z=0
\end{eqnarray}
We ignore the interaction here having in mind that it is reabsorbed in the quantum renormalization 
of mass.  The plane wave solution is $z(t,\vec{x}) \to u_{\lambda} e^{i\vec{k}\cdot\vec{x} -i\omega_{k\lambda} t}$, where
the spinor $u_{\lambda}$ follows from $(\vec{\sigma}\cdot\vec{B})u_{\lambda}=\lambda Bu_{\lambda}$, $\lambda=\pm 1$.
Hence
 \begin{equation}
\label{om}
 \omega_{k\lambda}=\Omega_{k} + \lambda {\cal B} \ .  
\end{equation}
Here $\Omega_{k}=\sqrt{m^2+k^2}$.
The standard canonical quantization gives
\begin{eqnarray}
z=\sum_{\vec{k}\lambda}\frac{1}{\sqrt{2\Omega_{k}}}
u_\lambda[\hat{a}_{k\lambda}e^{i\omega_{k\lambda} t - i\vec{k}\cdot\vec{x}} + \hat{b}_{k\lambda}^\dag e^{-i\omega_{k\lambda} t + i\vec{k}\cdot\vec{x}}]
\end{eqnarray}
where, $\hat{a}_{k\lambda}$ and $\hat{b}_{k\lambda}$ are bosonic creation and annihilation operators.
Importantly, $z$ is a usual spinor, $z \ne z^{\dag}$, and therefore one must introduce two types of bosons, $a$ and $b$, or spinons and antispinons.
This implies that at a given momentum there are four degrees of freedom, 
$(a,b)\times(\lambda=\pm 1)$~\cite{Senthil,SenthilA}.

\begin{figure}[ht]
 {\includegraphics[width=0.246\textwidth,clip]{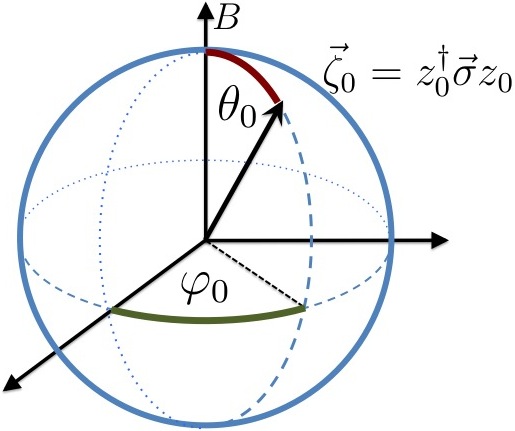}
}
 \caption{({\it Color online.
Orientation of the spinor condensate vector $\vec{\zeta}_0$ with respect to magnetic field.
})}
\label{F1}
\end{figure}

Obviously if ${\cal B} > m$ the dispersion (\ref{om}) becomes negative and this implies Bose condensation at $k=0$.
Energy density corresponding to (\ref{L}) is $E=\dot{z}^\dag \dot{z} + \nabla z^\dag\nabla z 
+ (m^2-{\cal B}^2)z^\dag z + \frac{\alpha}{2}(z^\dag z)^2$.
Minimization of the energy gives the spinor condensate and the classical energy~\cite{com2}
\begin{equation}
\label{cond}
z_0^\dag z_0 = \frac{({\cal B}^2-m^2)}{\alpha} \ , \ \ \ E_0= -\frac{({\cal B}^2-m^2)^2}{\alpha}.
\end{equation}
This expression has the following implications. First of all, $z_0^\dag z_0$ is a real number that effectively 
counts together the number of spinons and antispinons.
However, the condensate  $z_0^\dag z_0$ gives 
no information as to the relative contributions of 
spinons and antispinons.
The second implication, is that the ground state energy has no dependence on how the 
spin polarization vector  
$ \vec{\zeta}_0=z_0^\dag\vec{\sigma}z_0=(\sin\theta_0\cos\varphi_0,\sin\theta_0\sin\varphi_0,\cos\theta_0)$ 
is directed relative to the magnetic field (${\vec B} || {\hat z}$), 
see Fig.\ref{F1}.
%%%%%%%%%%%%%%%%%
%%%%%%%%%%%%%%%%%
The spinor condensate is
\begin{equation}
\label{z0}
z_0=A_0e^{i\gamma_0}\begin{pmatrix}
\cos\theta_0/2\\
e^{i\varphi_0}\sin\theta_0/2
\end{pmatrix},
\end{equation}
$A_0=\sqrt{({\cal B}^2-m^2)/\alpha}$.
Note, that the induced magnetization is still directed along $\vec B$, 
$\vec M=-\frac{\partial E_0}{\partial B}=4\mu_B\frac{({\cal B}^2-m^2)}{\alpha}{\vec {\cal B}}$.
The degeneracy with respect to direction of $ \vec{\zeta}_0$ implies that there are two 
Goldstone gapless excitations in the system 
corresponding to variation of angles $\theta_0$ and $\varphi_0$ in Fig.\ref{F1}.
We already pointed out that the total number of modes is four. So, in the spinor Bose condensate
phase we expect two gapless Goldstone modes and two gapped Higgs modes.
Note that the situation is very different from Bose condensation of magnons in a 
quantum antiferromagnet (condensation of a vector field). The staggered magnetization 
is always orthogonal to the external magnetic field. This results in one gapless Goldstone
mode and two gapped Higgs modes in the case of  magnon condensation, see e.g. 
Refs.~\cite{Sachdevphystoday,Kulik}.

To find explicitly the excitation modes in 
the Bose condensate phase we represent
\begin{equation}
\label{dz}
z=z_0+\delta z \ .
\end{equation}
The Euler-Lagrange equation for $\delta z$ reads
\begin{eqnarray}
\label{ELdz}
\delta\ddot{z} - \nabla^2\delta z - 2i(\vec{\sigma}\cdot\vec{{\cal B}
})\delta\dot{z} + \alpha(\delta z^\dag z_0)z_0 + \alpha(z_0^\dag\delta z)z_0 =0
\end{eqnarray}
A small mathematical complication is that this equation contains both $\delta z$ and
$\delta z^{\dag}$. Therefore, the solution is of the following form,
\begin{equation}
\label{dza}
\delta z=z_+e^{i\omega t - ikr} + z_-e^{-i\omega t +ikr}\ .
\end{equation}
After some algebra, presented in the supplementary material, we find the following modes
\begin{eqnarray}
\label{modes}
\omega_{1k}&=&\sqrt{{\cal B}^2+k^2}-{\cal B}\\
\omega_{2k}&=&\sqrt{3{\cal B}^2-m^2 + k^2 - \sqrt{(3{\cal B}^2-m^2)^2 + 4{\cal B}^2k^2}} \nonumber\\
\omega_{3k}&=&\sqrt{{\cal B}^2+k^2}+{\cal B}\nonumber\\
\omega_{4k}&=&\sqrt{3{\cal B}^2-m^2 + k^2 + \sqrt{(3{\cal B}^2-m^2)^2 + 4{\cal B}^2k^2}}\ .\nonumber
\end{eqnarray}
The first two modes are gapless (Goldstone), $\omega_{k=0}=0$, and the second two
modes are gapped (Higgs). Very surprisingly the first Goldstone mode has quadratic
dispersion at small $k$, $\omega_{1k} \approx \frac{k^2}{2{\cal B}}$, while the second
Goldstone mode has the usual linear dispersion, $\omega_{2k} \approx ck$, 
$c=\sqrt{\frac{{\cal B}^2-m^2}{3{\cal B}^2-m^2}}$.

\begin{figure*}[ht]
  \subfigure[]{\includegraphics[width=0.246\textwidth,clip]{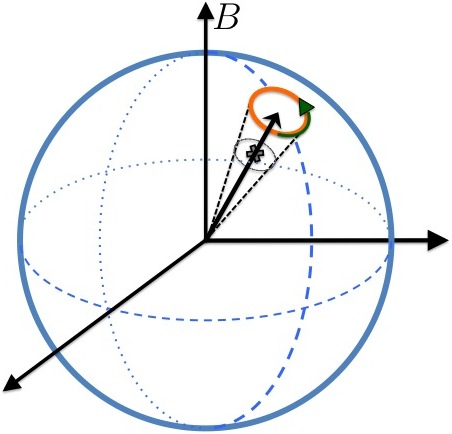}}
 \subfigure[]{\includegraphics[width=0.246\textwidth,clip]{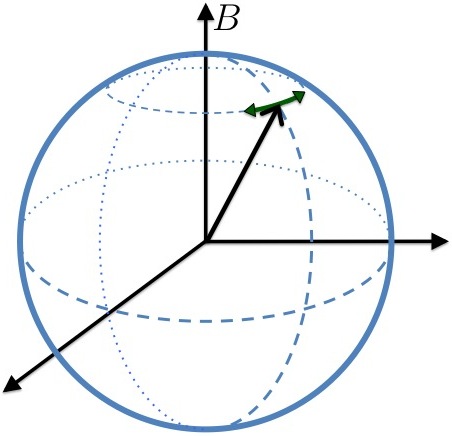}}
 \subfigure[]{\includegraphics[width=0.246\textwidth,clip]{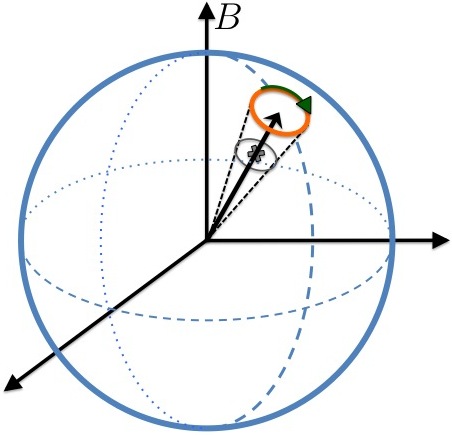}}
 \subfigure[]{\includegraphics[width=0.246\textwidth,clip]{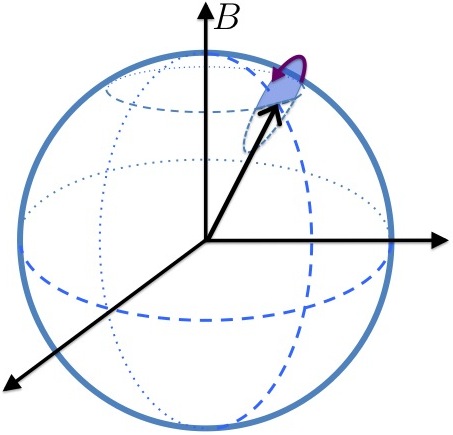}}
\caption{\it
Oscillations of the spin polarization vector $\vec{\zeta}$ with respect to the condensate vector $\vec{\zeta}_0$
in different excitation modes.  The solid black arrow represents an arbitrary $\vec{\zeta}_0$. 
(a) Goldstone mode with quadratic dispersion.  Anticlockwise rotation, the orange circle shows the path traced 
by the oscillations. 
(b) Goldstone mode with linear dispersion.
Linear oscillations in the direction perpendicular to the meridian.
(c) Higgs mode $|3\rangle$. Clockwise rotation, the orange circle shows the path traced 
by the oscillations. 
(d) Higgs mode $|4\rangle$. Elliptic oscillations  in the  plane perpendicular the meridian.
}
\label{F2}
\end{figure*}

In order to gain an intuitive/geometric understanding of the excitation modes, 
it is instructive to look at the variation of the spin expectation vector in these
modes, $\vec{\zeta}=z^{\dag}\vec{\sigma}z=\vec{\zeta}_0+\delta\vec{\zeta}$ where
$\delta\vec{\zeta}\approx z_0^{\dag}\vec{\sigma}\delta z + \delta z^{\dag}\vec{\sigma}z_0$.
Calculations presented in the supplementary material shows that for the Goldstone modes at small momenta $k\to0$, the variations are
\begin{eqnarray}
\delta\vec{\zeta}_1&\propto& \begin{pmatrix}
\cos\theta_0 \cos(\psi+\varphi_0)\\
\cos\theta_0 \sin(\psi +\varphi_0)\\
-\sin\theta_0 \cos\psi
\end{pmatrix}\nonumber \\
\delta\vec{\zeta}_2 &\propto& \begin{pmatrix}
-\sin\theta_0 \sin\varphi_0\\
\sin\theta_0\cos\varphi_0\\
0
\end{pmatrix}\sin\psi.
\end{eqnarray}
Here $\psi$ is the plane wave phase, $\psi=\omega t -{\vec k}\cdot{\vec r}$.
For the  Goldstone mode with quadratic dispersion, the variation $\delta\vec{\zeta}_1$ is shown in 
the panel (a) of Fig.~\ref{F2}. It represents an anticlockwise rotation of vector $\vec{\zeta}$ 
around the vacuum polarization vector $\vec{\zeta}_0$.
%%%%%%%%%%%%%%%%%%%
%%%%%%%%%%%%%%%%%%%%
The spin polarization variation $\delta\vec{\zeta}_2$ in  the  Goldstone mode with linear dispersion is shown 
in the panel (b) Fig.~\ref{F2}. The mode represents linear oscillations of vector $\vec{\zeta}$ around the vacuum
polarization vector $\vec{\zeta}_0$ in the direction perpendicular the meridian.

Both Goldstone modes satisfy the orthogonality condition
$\delta\vec{\zeta}_1\cdot\vec{\zeta}_0=\delta\vec{\zeta}_2\cdot\vec{\zeta}_0=0$.
The spin oscillations are in the plane orthogonal to the spin condensate vector.
Polarizations of the Goldstone modes are intuitively quite natural, a circular
polarization for the quadratic dispersion mode, similar to that in a ferromagnet and a
linear polarization in the linear dispersion mode, similar to that in an antiferromagnet.

Variations of the spin expectation vectors in Higgs modes are (see supplementary material).
\begin{eqnarray}
\delta\vec{\zeta}_3& \propto&\begin{pmatrix}
-\cos\theta_0 \cos(\psi-\varphi_0)\\
\cos\theta_0 \sin(\psi - \varphi_0)\\
\sin\theta_0 \cos\psi
\end{pmatrix}\nonumber \\ 
\delta\vec{\zeta}_4&\propto&\begin{pmatrix}
-\sin\theta_0 \sin\varphi_0\\
\sin\theta_0\cos\varphi_0\\
0
\end{pmatrix}\sin\psi + b_{4}\vec{\zeta}_0\cos\psi.
\end{eqnarray}
Where $b_4$, presented in the supplementary material, is a non-zero constant $b_4\to const\neq0$ at small momenta $k \to 0$.
The third mode represents a clockwise rotation of vector $\vec{\zeta}$ around the vacuum
polarization vector $\vec{\zeta}_0$, see panel (c) in Fig.~\ref{F2}.
Finally the fourth mode represents elliptic oscillations in the plane perpendicular to the meridian,
see panel (d) in Fig.~\ref{F2}.
The mode has both a longitudinal component (the component parallel to $\vec{\zeta}_0$) and a transverse component.

Instead of the external magnetic field a Quantum Phase Transition (QPT) can be driven by 
some external parameter g. In the case of magnon condensation in the spin
dimerized compound TlCuCl$_3$ such parameter is pressure~\cite{Tanaka,Ruegg04,Ruegg2014}.
In this case the mass squared term in the magnon Lagrangian  changes sign 
at the transition~\cite{Kulik}.  Now we consider a similar QPT for spinons
described by the Lagrangian (\ref{L}),  $m^2\propto g-g_c$. 
The formula $m^2\propto g-g_c$ assumes a continuous QPT, but in the end the continuity is 
not crucial, one can consider a weak discontinuous (first order) QPT too.
There are two following  crucial points which generally are independent of the kind of QPT,
(i) the nature of quasiparticles is the same (spinons)
on both sides of the QPT, (ii) the mass squared is negative, $m^2=-|m^2|$,
in the ordered phase.
Now we consider zero magnetic field.
Minimization of the energy gives the following spinor condensate in the ordered phase
\begin{equation}
\label{cond1}
z_0^\dag z_0 = \frac{|m^2|}{\alpha} \ .
\end{equation}
The Euler-Lagrange equation for the variation $\delta z$,
$\delta\ddot{z} - \nabla^2\delta z + \alpha(\delta z^\dag z_0)z_0 + \alpha(z_0^\dag\delta z)z_0 =0
$, can be easily solved using  (\ref{dza}). This gives three Goldstone excitations
and one gapped Higgs excitation
\begin{eqnarray}
\label{modes1}
\omega_1&=&\omega_2=\omega_3=k\nonumber\\
\omega_4&=&\sqrt{2|m^2| + k^2}\ .\nonumber
\end{eqnarray}
 
It is instructive to compare QPTs of spinons (condensation of the spinor field,
S=1/2) and QPTs of magnons (condensation of the vector field, S=1).
First, compare without magnetic field.
(i)In the case of magnons (S=1) there are three degenerate  gapped modes in the magnetically 
disordered phase. In the ordered phase there are two gapless Goldstone modes with linear
dispersion and one gapped Higgs mode (longitudinal magnon). 
(ii)In the case of spinons (S=1/2) there are four degenerate  gapped modes in the 
disordered phase. In the ordered phase there are three gapless Goldstone modes with linear
dispersion and one gapped Higgs mode.\\
Application of a sufficiently strong magnetic field in the disordered phase leads to 
Bose condensation, this is true for both magnons and spinons. However, excitations in the Bose
condensates are very different.
(i)In the case of Bose condensation of magnons (S=1) there is one gapless Goldstone mode with 
linear dispersion and two gapped Higgs modes. The linear Goldstone mode is a
generic property of conventional Bose condensation.
(ii)In the case of Bose condensation of spinons (S=1/2) there are two gapless Goldstone modes
and two gapped Higgs modes. One of the Goldstone modes has  quadratic dispersion and another
has linear dispersion. 

Existence of the quadratic Goldstone mode in the Bose condensate of spinons is the most
unconventional property. 
The first consequence of this property is that the condensate does not support superfluidity.
The Landau criterion of superfluidity~\cite{Landau} is not fulfilled. 
The second consequence of the quadratic mode is the unusual temperature dependence of the specific heat.
In a conventional Bose condensate the specific heat scales with temperature as $C\propto T^2$ in 2D,
and $C\propto T^3$ in 3D. At the same time in the S=1/2 Bose condensate
\begin{eqnarray}
\label{ch}
2D: \ \ \ C&=&\frac{1}{\pi}\zeta(2){\cal B} T\nonumber\\
3D: \ \ \ C&=&\frac{15}{8\sqrt{2}\pi^{3/2}}\zeta(5/2){\cal B}^{3/2}T^{3/2} \ .
\end{eqnarray}
Here $\zeta(x)$ is Riemann's zeta function.
The unconventional specific heat is a smoking gun for the unconventional Bose condensate.
The Mermin-Wagner theorem~\cite{mermin} is certainly valid for spinon condensates.
This implies that there is no true long range order at $T\ne 0$ in 2D case.
However, as usual the exponentially large correlation length does not influence the
power of temperature in the specific heat (\ref{ch}).

When applying the present analysis to the AF $\to$ VBS transition 
in a quantum magnet one has to remember that the deconfined description 
is valid only within a vicinity of the quantum critical point. There 
is no doubt that deep 
inside the VBS phase the quasiparticles are usual triplons and that deep 
inside the AF phase the quasipraticles are usual transverse magnons.
This implies that there is a spinon confinement length $\xi$ which 
depends on the detuning from the quantum critical point. 
It is possible that the gauge field~\cite{Senthil,SenthilA}, while does not
show up as a dynamic variable in (\ref{L}), still acts as a constraint 
contributing to formation of the confinement length.
The deconfined description, and in particular Eq.(\ref{ch}), is valid
if the spinon thermal wavelength is smaller than the confinement scale,
$\lambda_T \sim 1/\sqrt{{\cal B} T} \ll \xi$. 
%Evolution of the Bose condensate specific heat with detuning from the quantum 
%critical point depends

In conclusion, we have considered Bose condensation of particles with spin S=1/2.
The Bose statistics for the half integer spin particles implies violation of the spin statistics
theorem. While we believe that our results are generic, specifically we consider the Bose 
condensation of spinons driven by a magnetic field.
The most surprising are the gapless Goldstone excitations in the spinon condensate.
There are two such excitations, one has a linear in momentum dispersion and 
another has a quadratic dispersion. The quadratic dispersion implies that the
condensate does not support superfluidity. This also implies an unconventional
temperature behaviour of the specific heat. 
This is  a smoking gun criterion/behaviour for identification of the unconventional Bose 
condensation.
Observation of such behaviour in an experiment or in a numerical simulation would be an 
unambiguous indication for the Bose condensation of spinons.

We thank A. W. Sandvik for discussions.


\begin{thebibliography}{99}
\bibitem{com1} The Cooper pairing of particles with half integer spin leads to superfluidity, 
say liquid $^3$He at mK temperatures. However, this is the Bose condensation of pairs and a pair
has an integer spin.
\bibitem{com2}
We treat the energy in a simple mean-field approximation.
There is no doubt that quantum fluctuations/corrections significantly modify
Eqs.(\ref{cond}) and some other Eqs., especially in the 2D case.
However, the goal of the present work is to find properties of quasiparticles/excitations.
These properties are dictated by symmetries and hence they are not influenced by the 
dynamic approximations made. This point is valid for a usual quantum antiferromagnet~\cite{Kulik}
(condensation of a vector field) and it is equally valid for condensation of spinons.
\end{thebibliography}
\end{document}

\chapter*{Supplementary Material}
\section*{Part A}
\textit{In this section we show how the Euler-Lagrange equation (8)
generates the Bose condensate excitation modes (10).}\\
It is convenient to define the following time independent spinors
\begin{equation}
\tag{A.1}
u=\begin{pmatrix}
\cos\frac{\theta_0}{2}\\
\sin\frac{\theta_0}{2}e^{i\varphi_0}
\end{pmatrix} \ , \ \ \ 
\bar{u}= \begin{pmatrix}
-\sin\frac{\theta_0}{2}\\
\cos\frac{\theta_0}{2}e^{i\varphi_0}
\end{pmatrix}\ .
\end{equation}
Hence the Bose condensate field (6) is $z_0=A_0 u$.
Substitution of ansatz (9) in the Euler-Lagrange equation (8)
results in the following algebraic equation for the ``amplitude'' spinors
$z_{+}$ and $z_{-}$
\begin{align*}
\tag{A.2}
(-\omega^2+k^2)z_{+} + 2\omega (\vec{\sigma}\cdot\vec{{\cal B}})z_{+} + ({\cal B}^2-m^2)(z_{-}^\dag u + u^\dag z_{+})u&=0\\
(-\omega^2+k^2)z_{-} - 2\omega (\vec{\sigma}\cdot\vec{{\cal B}})z_{-} + ({\cal B}^2-m^2)(z_{+}^\dag u + u^\dag z_{-})u&=0.
\end{align*}
Projecting each equation (A.2), onto both $u$ and $\bar{u}$, and using the following definitions;
\begin{align*}
\tag{A.3}
u^\dag z_{+} &= a_1            &&a_1 + a_2^*= a_+\\
u^\dag z_{-} &= a_2              &&a_1 - a_2^*=a_-\\
\bar{u}^\dag z_{+} &= b_1        &&b_1+b_2^*\,=b_+\\
\bar{u}^\dag z_{-} &= b_2  &&b_1-b_2^*\,=b_-\\
\end{align*}
we get the following matrix equation for the amplitudes
$a_+$, $a_-$, $b_+$, $b_-$,
 \begin{align*}
\tag{A.4}\begin{pmatrix} 
k^2 - \omega^2 + 2({\cal B}^2-m^2) & 2\omega {\cal B}\cos\theta_0& 0 & -2\omega {\cal B}\sin\theta_0 \\
2\omega {\cal B}\cos\theta_0& k^2 - \omega^2& -2\omega {\cal B}\sin\theta_0&0\\
0& -2\omega {\cal B}\sin\theta_0 & k^2 - \omega^2& -2\omega {\cal B}\cos\theta_0\\
-2\omega {\cal B}\sin\theta_0  & 0 &- 2\omega{\cal B}\cos\theta_0&k^2 - \omega^2\\
\end{pmatrix}
\begin{pmatrix}
a_+\\
a_-\\
b_+\\
b_-\\
\end{pmatrix}
=0.
\end{align*}
Frequencies of normal modes (10) immediately follow from this matrix 
equation.
Amplitude vectors  $( a_+, a_-, b_+, b_-)$,
corresponding to the normal modes are of the following form
\begin{align*}
&|1\rangle \propto 
\begin{pmatrix}
0\\
\sin\theta_0\\
1\\
\cos\theta_0\\
\end{pmatrix}; \ \ \ \
|2\rangle \propto 
\begin{pmatrix}
b_2\\
-\cos\theta_0\\
0\\
\sin\theta_0\\
\end{pmatrix}; \ \ \ \
|3\rangle \propto
\begin{pmatrix}
0\\
\sin\theta_0\\
-1\\
\cos\theta_0\\
\end{pmatrix}; \ \ \ \
|4\rangle \propto
\begin{pmatrix}
b_4\\
-\cos\theta_0\\
0\\
\sin\theta_0\\
\end{pmatrix}\\
\tag{A.5}
&b_2=\frac{\sqrt{(3{\cal B}^2-m^2)^2+4{\cal B}^2k^2}-(3{\cal B}^2-m^2)}
{2{\cal B}\omega_2}\nonumber\\
&b_4=\frac{-\sqrt{(3{\cal B}^2-m^2)^2+4{\cal B}^2k^2}-(3{\cal B}^2-m^2)}
{2{\cal B}\omega_4}\nonumber
\end{align*}
At small momenta, $k\to 0$, the coefficients behave as
$b_2 \to 0$, $b_4 \to const \ne 0$.
Note that  there is no naive orthogonality of ``eigenvectors'', in particular 
$\langle 2|4\rangle \ne 0$.
A similar nonorthogonality is typical for classical (non-quantum)
coupled oscillators in magnetic field.
In spite of the nonorthogonality, after an appropriate canonical transformation
the Hamiltonian is transformed to the sum of independent Hamiltonians
of the normal modes.

\newpage
\section*{Part B}
\textit{
In this section we calculate spin vibrations (11),(12)
corresponding to different excitation modes.}\\
Using Eqs. (A.1) and (A.3) we find
\begin{align*}
\tag{B.1}
z_+&\propto \begin{pmatrix}
[(a_{+}+a_{-})\cos\frac{\theta_0}{2} - (b_{+}+b_{-})\sin\frac{\theta_0}{2}]\\
[(a_{+}+a_{-})\sin\frac{\theta_0}{2} + (b_{+}+b_{-})\cos\frac{\theta_0}{2}]e^{i\varphi_0}\\
\end{pmatrix}\\
z_-&\propto \begin{pmatrix}
[(a_{+}-a_{-})\cos\frac{\theta_0}{2} - (b_{+}-b_{-})\sin\frac{\theta_0}{2}]\\
[(a_{+}-a_{-})\sin\frac{\theta_0}{2} + (b_{+}-b_{-})\cos\frac{\theta_0}{2}]e^{i\varphi_0}
\end{pmatrix}\\
\end{align*}
From here, using
\begin{align*}
\tag{B.2}
\delta\vec{\zeta}&=\delta z^{\dag}\vec{\sigma}z_0 + z_0^{\dag}\vec{\sigma}\delta z,
\end{align*}
together with Eqs.(9) and(A.5) we find explicit formulas for
 the spin vibration vectors 
$\delta {\vec \zeta}$ presented in Eqs. (11) and (12).